\documentclass{article} 

\usepackage{hyperref}
\usepackage{url}

\usepackage[accepted]{icml2015}

\usepackage{algorithm,algorithmic,color}
\usepackage{float}
\usepackage{amssymb}
\usepackage{amsmath}
\usepackage{amsthm}
\usepackage{graphicx}
\usepackage{color}
\usepackage{stmaryrd}
\usepackage{placeins}

\definecolor{darkblue}{rgb}{0,0.05,0.35}
\definecolor{darkgreen}{rgb}{0,0.6,0}

\newcommand{\rev}[1]{\overset{{}_{\shortleftarrow}}{#1}}

\newcommand{\E}{\mathbb{E}}
\newcommand{\I}{\mathbb{I}}

\newcommand{\qT}{q_{\theta}}
\newcommand{\rT}{r_{\theta}}

\newcommand{\eqnr}{\addtocounter{equation}{1}\tag{\theequation}}

\icmltitlerunning{Markov Chain Monte Carlo and Variational Inference: Bridging the Gap}

\begin{document}

\twocolumn[
\icmltitle{Markov Chain Monte Carlo and Variational Inference:\\Bridging the Gap}

\icmlauthor{Tim Salimans}{tim@algoritmica.nl}
\icmladdress{Algoritmica}
\icmlauthor{Diederik P. Kingma and Max Welling}{[d.p.kingma,m.welling]@uva.nl}
\icmladdress{University of Amsterdam}

\icmlkeywords{}

\vskip 0.3in
]

\begin{abstract}
Recent advances in stochastic gradient variational inference have made it possible to perform variational Bayesian inference with posterior approximations containing auxiliary random variables. This enables us to explore a new synthesis of variational inference and Monte Carlo methods where we incorporate one or more steps of MCMC into our variational approximation. By doing so we obtain a rich class of inference algorithms bridging the gap between variational methods and MCMC, and offering the best of both worlds: fast posterior approximation through the maximization of an explicit objective, with the option of trading off additional computation for additional accuracy. We describe the theoretical foundations that make this possible and show some promising first results.
\end{abstract}

\section{MCMC and Variational Inference}
\label{sec:stochvb}
Bayesian analysis gives us a very simple recipe for learning from data: given a set of unknown parameters or latent variables $z$ that are of interest, we specify a prior distribution $p(z)$ quantifying what we know about $z$ before observing any data. Then we quantify how the observed data $x$ relates to $z$ by specifying a likelihood function $p(x|z)$. Finally, we apply Bayes' rule $p(z|x) = p(z)p(x|z)/\int p(z)p(x|z) dz$ to give the posterior distribution, which quantifies what we know about $z$ after seeing the data.

Although this recipe is very simple conceptually, the implied computation is often intractable. We therefore need to resort to approximation methods in order to perform Bayesian inference in practice. The two most popular approximation methods for this purpose are variational inference and Markov Chain Monte Carlo (MCMC). The former has the advantage of maximizing an explicit objective, and being faster in most cases. The latter has the advantage of being nonparametric and asymptotically exact. Here, we show how both methods can be combined in order to get the best of both worlds. 

\subsection{Variational Inference}
\textit{Variational inference} casts Bayesian inference as an optimization problem where we introduce a parameterized posterior approximation $\qT(z|x)$ which is fit to the posterior distribution by choosing its parameters $\theta$ to maximize a lower bound $\mathcal{L}$ on the marginal likelihood:
\begin{align*}
\log p(x) 
&\geq \log p(x) -  D_{KL}(\qT(z|x)||p(z|x))\eqnr\\
&= \E_{\qT(z|x)}[\log p(x,z) - \log \qT(z|x)] = \mathcal{L}.
\label{eq:lowerbound}
\eqnr\end{align*}
Since $\log p(x)$ is independent of $\theta$, maximizing the bound $\mathcal{L}$ w.r.t. $\theta$ will minimize the KL-divergence $D_{KL}(\qT(z|x)||p(z|x))$. The bound above is tight at $D_{KL}(\qT(z|x)||p(z|x)) = 0$, when the approximation $\qT(z|x)$ perfectly matches $p(z|x)$.

\subsection{MCMC and Auxiliary Variables}
A popular alternative to variational inference is the method of \textit{Markov Chain Monte Carlo} (MCMC). Like variational inference, MCMC starts by taking a random draw $z_{0}$ from some initial distribution $q(z_{0})$ or $q(z_{0}|x)$. Rather than optimizing this distribution, however, MCMC methods subsequently apply a \textit{stochastic transition operator} to the random draw $z_{0}$:
\[ z_{t} \sim q(z_{t}|z_{t-1},x). \]
By judiciously choosing the transition operator $q(z_{t}|z_{t-1},x)$ and iteratively applying it many times, the outcome of this procedure, $z_{T}$, will be a random variable that converges in distribution to the exact posterior $p(z|x)$. The advantage of MCMC is that the samples it gives us can approximate the exact posterior arbitrarily well if we are willing to apply the stochastic transition operator a sufficient number of times. The downside of MCMC is that in practice we do not know how many times is sufficient, and getting a good approximation using MCMC can take a very long time.

The central idea of this paper is that we can interpret the stochastic Markov chain $q(z|x)=q(z_{0}|x)\prod_{t=1}^{T}q(z_{t}|z_{t-1},x)$ as a variational approximation in an expanded space by considering $y = z_{0},z_{1},\ldots,z_{t-1}$ to be a set of \textit{auxiliary random variables}. Integrating these auxiliary random variables into the variational lower bound \eqref{eq:lowerbound}, we obtain
\begin{align*}
&\mathcal{L}_{\text{aux}} \eqnr\label{eq:auxlowerbound}\\
&= \E_{q(y,z_T|x)}[\log[p(x,z_T)r(y|x,z_T)] - \log q(y,z_T|x)] \\
&= \mathcal{L} - \E_{q(z_T|x)}\{D_{KL}[q(y|z_T,x)||r(y|z_T,x)]\}\\
&\leq \mathcal{L} \hspace{0.1cm} \leq \hspace{0.1cm} \log[p(x)],
\end{align*}
where $r(y|x,z_T)$ is an auxiliary inference distribution which we are free to choose, and our marginal posterior approximation is given by $q(z_T|x) = \int q(y,z_T|x)\text{d}y$. The marginal approximation $q(z_T|x)$ is now a mixture of distributions of the form $q(z_T|x,y)$. Since this is a very rich class of distributions, auxiliary variables may be used to obtain a closer fit to the exact posterior \cite{salimans2013fixed}. The choice $r(y|x,z_T) = q(y|x,z_T)$ would be optimal, but again often intractable to compute; in practice, good results can be obtained by specifying a $r(y|x,z_T)$ that can approximate $q(y|x,z_{T})$ to a reasonable degree. One way this can be achieved is by specifying $r(y|x,z_T)$ to be of some flexible parametric form, and optimizing the lower bound over the parameters of this distribution. In this paper we consider the special case where the auxiliary inference distribution also has a Markov structure just like the posterior approximation: $r(z_0,\ldots,z_{t-1}|x,z_T) = \prod_{t=1}^T r_t(z_{t-1}|x,z_t)$, in which case the variational lower bound can be rewritten as
\begin{align*}
\label{eq:auxbound}\eqnr
\log p(x) & \geq \E_q[ \log p(x,z_T) - \log q(z_0,\ldots,z_T|x) \\
&+ \log r(z_0,\ldots,z_{t-1}|x,z_T) ] \nonumber\\
&= \E_q \big[ \log[p(x,z_T)/q(z_0|x)] \\
&+ \sum_{t=1}^T\log[r_t(z_{t-1}|x,z_t)/q_t(z_t|x,z_{t-1})] \big].
\end{align*}
where the subscript $t$ in $q_{t}$ and $r_{t}$ highlights the possibility of using different transition operators $q_{t}$ and inverse models $r_{t}$ at different points in the Markov chain. By specifying these $q_{t}$ and $r_{t}$ in some flexible parametric form, we can then optimize the value of \eqref{eq:auxbound} in order to get a good approximation to the true posterior distribution.

\section{Optimizing the lower bound}
For most choices of the transition operators $q_{t}$ and inverse models $r_{t}$, the auxiliary variational lower bound \eqref{eq:auxbound} cannot be calculated analytically. However, if we can at least sample from the transitions $q_{t}$, and evaluate the inverse models $r_{t}$ at those samples, we can still approximate the variational lower bound without bias using the following algorithm:
\begin{algorithm}[H]
\caption{MCMC lower bound estimate}
\label{algo:mcmclb}
\begin{algorithmic}
\REQUIRE Model with joint distribution $p(x,z)$ and a desired but intractable posterior $p(z|x)$
\REQUIRE Number of iterations $T$
\REQUIRE Transition operator(s) $q_t(z_t|x,z_{t-1})$
\REQUIRE Inverse model(s) $r_t(z_{t-1}|x,z_t)$
\STATE Draw an initial random variable $z_0 \sim q(z_0|x)$
\STATE Initialize the lower bound estimate as\\ $L = \log p(x,z_0) - \log q(z_0|x)$
\FOR{$t = 1:T$}
\STATE Perform random transition $z_t \sim q_t(z_t|x,z_{t-1})$
\STATE Calculate the ratio $\alpha_t = \frac{p(x,z_t)r_t(z_{t-1}|x,z_t)}{p(x,z_{t-1})q_t(z_t|x,z_{t-1})}$
\STATE Update the lower bound $L = L + \log[\alpha_{t}]$
\ENDFOR
\RETURN the unbiased lower bound estimate $L$
\end{algorithmic}
\end{algorithm}
The key insight behind the recent work in \textit{stochastic gradient variational inference} is that if all the individual steps of an algorithm like this are differentiable in the parameters of $q$ and $r$, which we denote by $\theta$, then so is the algorithm's output $L$. Since $L$ is an unbiased estimate of the variational lower bound, its derivative is then an unbiased estimate of the derivative of the lower bound, which can be used in a stochastic optimization algorithm.

Obtaining gradients of the Monte Carlo estimate of Algorithm~\ref{algo:mcmclb} requires the application of the chain rule through the random sampling of the transition operators $q_t(z_t|x,z_{t-1})$. This can in many cases be realised by drawing from these operators in two steps: In the first step we draw a set of \textit{primitive random variables} $u_t$ from a fixed distribution $p(u_t)$, and we then transform those as $z_t = g_{\theta}(u_t,x)$ with a transformation $g_{\theta}()$ chosen in such a way that $z_t$ follows the distribution $q_t(z_t|x,z_{t-1})$. If this is the case we can apply backpropagation, differentiating through the sampling function to obtain unbiased stochastic estimates of the gradient of the lower bound objective with respect to $\theta$ \cite{salimans2013fixed,kingma2013auto,rezende2014stochastic}. An alternative solution, which we do not consider here, would be to approximate the gradient of the lower bound using Monte Carlo directly \cite{paisley2012variational,ranganath2013black,mnih2014neural}.



Once we have obtained a stochastic estimate of the gradient of \eqref{eq:lowerbound} with respect to $\theta$, we can use this estimate in a stochastic gradient-based optimization algorithm for fitting our approximation to the true posterior $p(z|x)$. We do this using the following algorithm:

\begin{algorithm}[H]
\caption{Markov Chain Variational Inference (MCVI)}
\label{algo:mcvi}
\begin{algorithmic}
\REQUIRE Forward Markov model $q_{\theta}(z)$ and backward Markov model $r_{\theta}(z_{0},\ldots,z_{t-1}|z_{T})$
\REQUIRE Parameters $\theta$
\REQUIRE Stochastic estimate $L(\theta)$ of the variational lower bound $\mathcal{L}_{\text{aux}}(\theta)$ from Algorithm~\ref{algo:mcmclb}
\WHILE{not converged}
\STATE Obtain unbiased stochastic estimate $\hat{g}$ with $E_{q}[\hat{g}] = \nabla_{\theta}\mathcal{L}_{\text{aux}}(\theta)$ by differentiating $L(\theta)$
\STATE Update the parameters $\theta$ using gradient $\hat{g}$ and a stochastic optimization algorithm
\ENDWHILE
\RETURN final optimized variational parameters $\theta$
\end{algorithmic}
\end{algorithm}

\subsection{Example: bivariate Gaussian}
\label{sec:gauss_gibbs}
As a first example we look at sampling from the bivariate Gaussian distribution defined by
\[
p(z^{1},z^{2}) \propto \exp\left[-\frac{1}{2\sigma_{1}^{2}}(z^{1}-z^{2})^{2}-\frac{1}{2\sigma_{2}^{2}}(z^{1}+z^{2})^{2}\right].
\]
We consider two MCMC methods that update the univariate $z^{1},z^{2}$ in turn. The first method is Gibbs sampling, which samples from the Gaussian full conditional distributions $p(z^{i}|z^{-i}) = N(\mu_{i},\sigma^{2}_{i})$. The second method is the over-relaxation method of \cite{adler1981over}, which instead updates the univariate $z^{i}$ using $q(z^{i}_{t}|z_{t-1}) = N[\mu_{i}+\alpha(z^{i}_{t-1}-\mu_{i}),\sigma^{2}_{i}(1-\alpha^{2})]$. For $\alpha=0$ the two methods are equivalent, but for other values of $\alpha$ the over-relaxation method may mix more quickly than Gibbs sampling. To test this we calculate the variational lower bound for this MCMC algorithm, and maximize with respect to $\alpha$ to find the most effective transition operator.

\begin{figure}[t]
\centering
\includegraphics[width=0.45\textwidth]{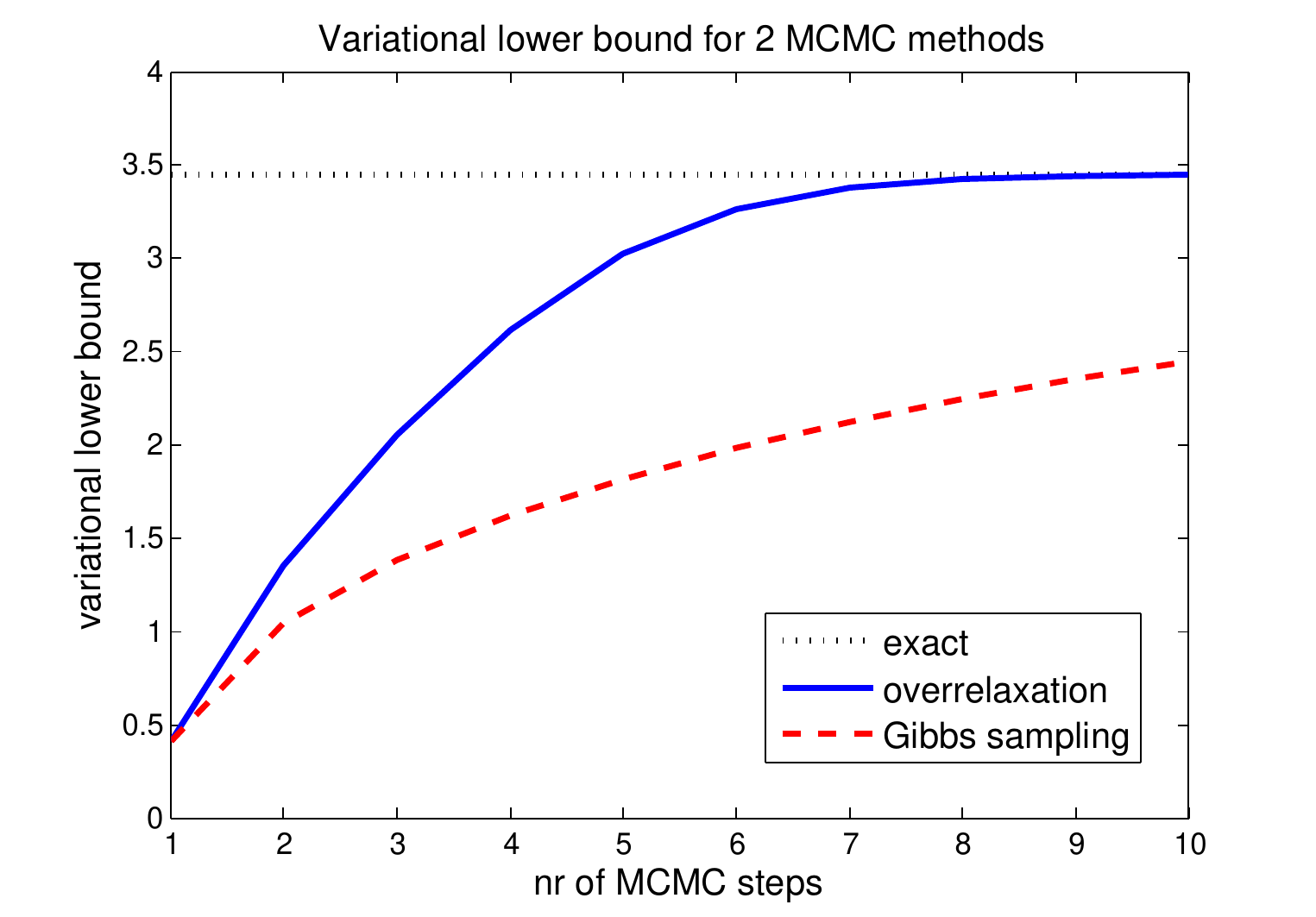}
\caption{The log marginal likelihood lower bound for a bivariate Gaussian target and an MCMC variational approximaton, using Gibbs sampling or Adler's overrelaxation.} 
\label{fig:gaussian}
\end{figure}

For the inverse model $r(z_{t-1}|z_{t})$ we use Gaussians with mean parameter linear in $z_{t}$ and variance independent of $z_{t-1}$ . For this particular case this specification allows us to recover the $q(z_{t-1}|z_{t})$ distribution exactly. We use $\sigma_{1}=1,\sigma_{2}=10$ in our exact posterior, and we initialize the Markov chain at $(-10,-10)$, with addition of infinitesimal noise (variance of $10^{-10}$). Figure \ref{fig:gaussian} shows the lower bound for both MCMC methods: over-relaxation with an optimal $\alpha$ of $-0.76$ clearly recovers the exact posterior much more quickly than plain Gibbs sampling. The fact that optimization of the variational lower bound allows us to improve upon standard methods like Gibbs sampling is promising for more challenging applications. 

\FloatBarrier

\section{Hamiltonian variational inference}
One of the most efficient and widely applicable MCMC methods is \textit{Hamiltonian Monte Carlo} (HMC) \cite{neal2011mcmc}. HMC is an MCMC method for approximating continuous distributions $p(z|x)$ where the space of unknown variables is expanded to include a set of auxiliary variables $v$ with the same dimension as $z$. These auxiliary variables are initialized with a random draw from a distribution $v_{t}' \sim q(v_{t}'|x,z_{t-1})$, after which the method simulates the dynamics corresponding to the Hamiltonian $H(v,z) = 0.5v^{T}M^{-1}v - \log p(x,z)$, where $z$ and $v$ are iteratively updated using the \textit{leapfrog integrator}, see \cite{neal2011mcmc}.

Hamiltonian dynamics of this form is a very effective way of exploring the posterior distribution $p(z|x)$ because the dynamics is guided by the gradient of the exact log posterior, and random walks are suppressed by the auxiliary variables $v$, which are also called \textit{momentum variables}. Furthermore, the transition from $v_{t}',z_{t-1}$ to $v_{t},z_{t}$ in HMC is deterministic, invertible and volume preserving, which means that we have
\begin{align*}
q(v_t,z_t|z_{t-1},x) = q(v_t,z_t,z_{t-1}|x)/q(z_{t-1}|x)\\
= q(v_{t}',z_{t-1}|x)/q(z_{t-1}|x) = q(v_{t}'|z_{t-1},x)
\end{align*}
and similarly $r(v_{t}',z_{t-1}|z_{t},x) = r(v_{t}|z_{t},x)$, with $z_{t},v_{t}$ the output of the Hamiltonian dynamics.

Using this choice of transition operator $q_{t}(v_t,z_t|z_{t-1},x)$ and inverse model $r_{t}(v_{t}',z_{t-1}|z_{t},x)$ we obtain the following algorithm for stochastically approximating the log marginal likelihood lower bound:
\begin{algorithm}[H]
\caption{Hamiltonian variational inference (HVI)}
\label{algo:hmcvb}
\begin{algorithmic}
\REQUIRE Unnormalized log posterior $\log p(x,z)$
\REQUIRE Number of iterations $T$
\REQUIRE Momentum initialization distribution(s) $q_{t}(v_{t}'|z_{t-1},x)$ and inverse model(s) $r_{t}(v_{t}|z_{t},x)$
\REQUIRE HMC stepsize and mass matrix $\epsilon,M$
\STATE Draw an initial random variable $z_{0} \sim q(z_{0}|x)$
\STATE Init. lower bound $L = \log[p(x,z_{0})] - \log[q(z_{0}|x)]$
\FOR{$t = 1:T$}
\STATE Draw initial momentum $v_{t}' \sim q_{t}(v_{t}'|x,z_{t-1})$
\STATE Set $z_{t},v_{t} = \text{Hamiltonian\_Dynamics}(z_{t-1},v_{t}')$ 
\STATE Calculate the ratio $\alpha_{t} = \frac{p(x,z_{t})r_{t}(v_{t}|x,z_{t})}{p(x,z_{t-1})q_{t}(v_{t}'|x,z_{t-1})}$
\STATE Update the lower bound $L = L + \log[\alpha_{t}]$
\ENDFOR
\RETURN lower bound $L$, approx. posterior draw $z_{T}$
\end{algorithmic}
\end{algorithm}
Here we omit the Metropolis-Hastings step that is typically used with Hamiltonian Monte Carlo. Section~\ref{sec:detail} discusses how such as step could be integrated into Algorithm~\ref{algo:hmcvb}.

We fit the variational approximation to the true posterior distribution by stochastically maximizing the lower bound with respect to $q$,$r$ and the parameters (stepsize and mass matrix) of the Hamiltonian dynamics using Algorithm~\ref{algo:mcvi}. We call this version of the algorithm \textit{Hamiltonian Variational Inference} (HVI).
After running the algorithm to convergence, we then have an optimized approximation $q(z|x)$ of the posterior distribution. Because our approximation automatically adapts to the local shape of the exact posterior, this approximation will often be better than a variational approximation with a fixed functional form, provided our model for $r_{t}(v_{t}|x,z_{t})$ is flexible enough. 

In addition to improving the quality of our approximation, we find that adding HMC steps to a variational approximation often reduces the variance in our stochastic gradient estimates, thereby speeding up the optimization. The downside of using this algorithm is that its computational cost per iteration is higher than when using an approximate $q(z|x)$ of a fixed form, mainly owing to the need of calculating additional derivatives of $\log p(x,z)$. These derivatives may also be difficult to derive by hand, so it is advisable to use an automatic differentiation package such as Theano \cite{bastien2012theano}. As a rule of thumb, using the Hamiltonian variational approximation with $m$ MCMC steps and $k$ leapfrog steps is about $mk$ times as expensive per iteration as when using a fixed form approximation. This may be offset by reducing the number of iterations, and in practice we find that adding a single MCMC step to a fixed-form approximation often speeds up the convergence of the lower bound optimization in wallclock time. The scaling of the computational demands in the dimensionality of $z$ is the same for both Hamiltonian variational approximation and fixed form variational approximation, and depends on the structure of $p(x,z)$.

Compared to regular Hamiltonian Monte Carlo, Algorithm~\ref{algo:hmcvb} has a number of advantages: The samples drawn from $q(z|x)$ are independent, the parameters of the Hamiltonian dynamics $(M,\epsilon)$ are automatically tuned, and we may choose to omit the Metropolis-Hastings step so as not to reject any of the proposed transitions. Furthermore, we optimize a lower bound on the log marginal likelihood, and we can assess the approximation quality using the techniques discussed in \cite{salimans2013fixed}. By finding a good initial distribution $q(z_{0})$, we may also speed up convergence to the true posterior and get a good posterior approximation using only a very short Markov chain, rather than relying on asymptotic theory.
\FloatBarrier

\subsection{Example: A beta-binomial model for overdispersion}
To demonstrate our Hamiltonian variational approximation algorithm we use an example from \cite{albert:09}, which considers the problem of estimating the rates of death from stomach cancer for the largest cities in Missouri. The data is available from the R package LearnBayes. It consists of 20 pairs $(n_{j},x_{j})$ where $n_{j}$ contains the number of individuals that were at risk for cancer in city $j$, and $x_{j}$ is the number of cancer deaths that occurred in that city. The counts $x_{j}$ are overdispersed compared to what one could expect under a binomial model with constant probability, so \cite{albert:09} assumes a beta-binomial model with a two dimensional parameter vector $z$. The low dimensionality of this problem allows us to easily visualize the results.

We use a variational approximation containing a single HMC step so that we can easily integrate out the 2 momentum variables numerically for calculating the exact KL-divergence of our approximation and to visualize our results. We choose $\qT(z_{0}),\qT(v_{1}'|z_{0}),\rT(v_{1}|z_{1})$ to all be multivariate Gaussian distributions with diagonal covariance matrix. The mass matrix $M$ is also diagonal. The means of $\qT(v_{1}'|z_{0})$ and $\rT(v_{1}|z_{1})$ are defined as linear functions in $z$ and $\nabla_{z}\log p(x,z)$, with adjustable coefficients. The covariance matrices are not made to depend on $z$, and the approximation is run using different numbers of leapfrog steps in the Hamiltonian dynamics. 

\begin{figure}[t]
\centering
\includegraphics[width=0.45\textwidth]{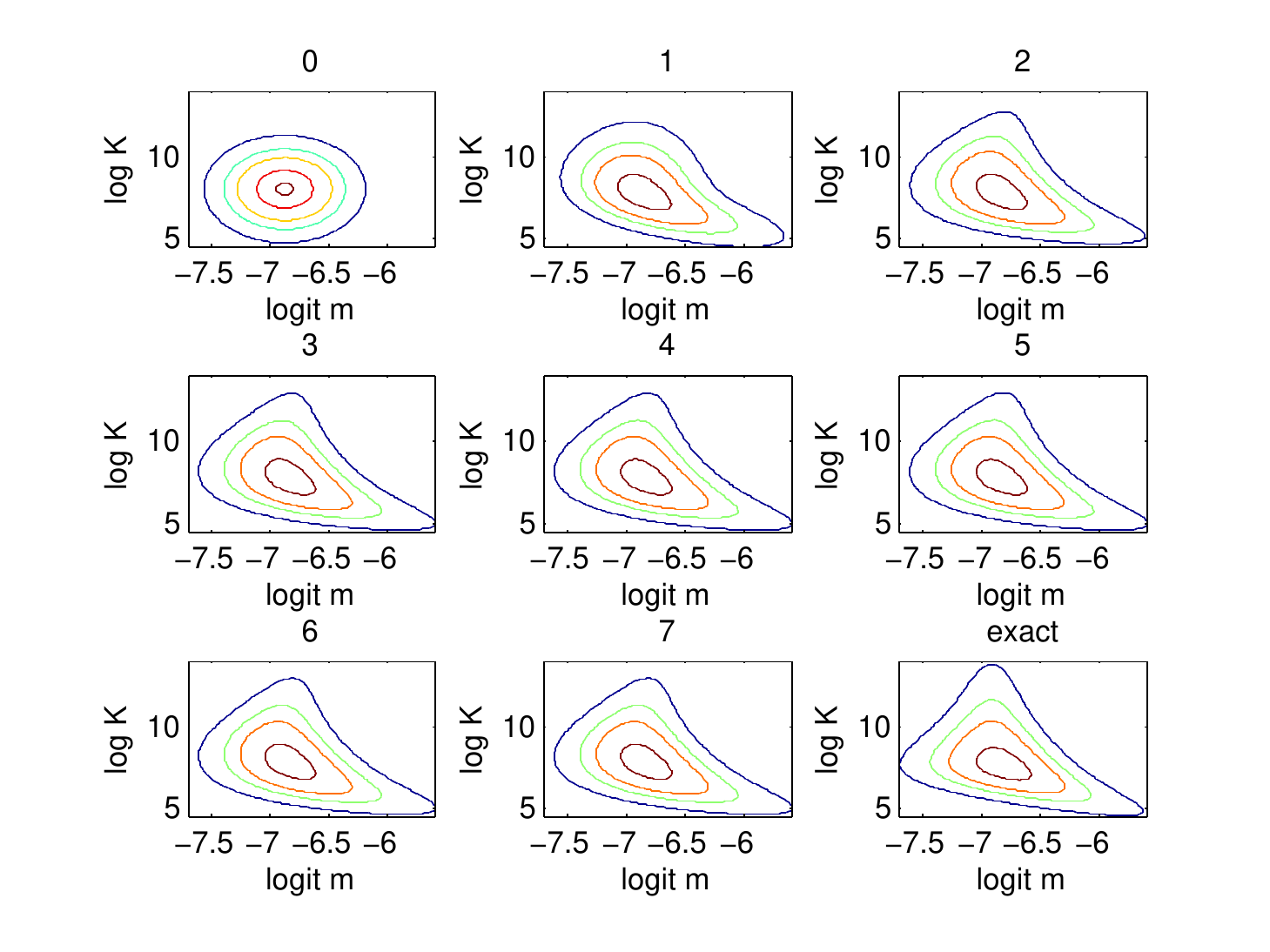}
\caption{Approximate posteriors for a varying number of leapfrog steps. Exact posterior at bottom right.}
\label{fig:leapfrog}
\end{figure}

\begin{figure}[htb]
\centering
\includegraphics[width=0.45\textwidth]{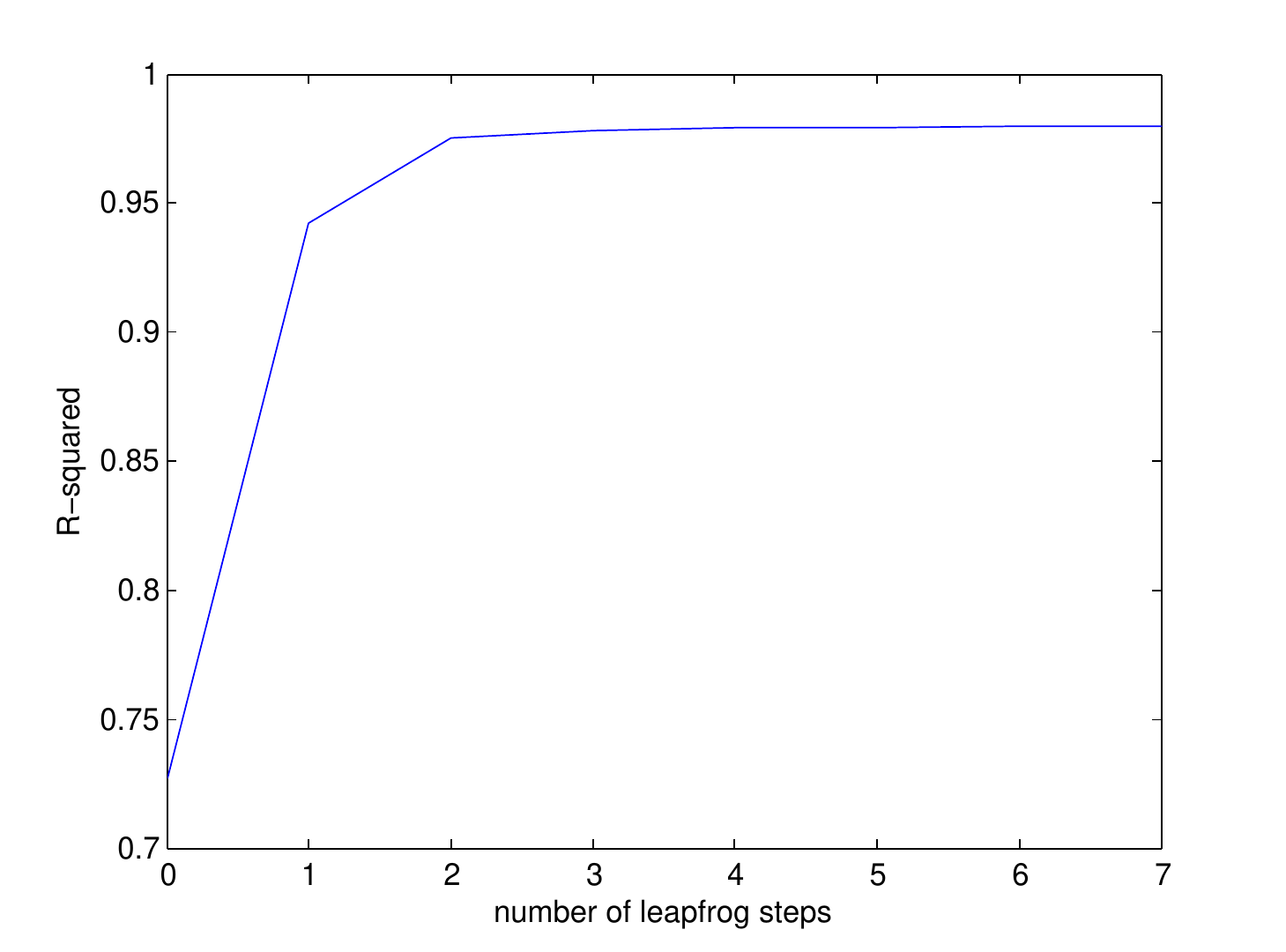}
\caption{R-squared accuracy measure \cite{salimans2013fixed} for approximate posteriors using a varying number of leapfrog steps.}
\label{fig:r2}
\end{figure}

As can be seen from Figures~\ref{fig:leapfrog} and~\ref{fig:r2}, the Hamiltonian dynamics indeed helps us improve the posterior approximation. Most of the benefit is realized in the first two leapfrog iterations. Of course, more iterations may still prove useful for different problems and different specifications of $\qT(z_{0}),\qT(v_{1}'|z_{0}),\rT(v_{1}|z_{1})$, and additional MCMC steps may also help. Adjusting only the means of $\qT(v_{1}'|z_{0})$ and $\rT(v_{1}|z_{1})$ based on the gradient of the log posterior is a simple specification that achieves good results. We find that even simpler parameterizations still do quite well, by finding a solution where the variance of $\qT(v_{1}'|z_{0})$ is larger than that of $\rT(v_{1}|z_{1})$, and the variance of $\qT(z_{0})$ is smaller than that of $p(v|z)$: The Hamiltonian dynamics then effectively transfers entropy from $v$ to $z$, resulting in an improved lower bound.


\subsection{Example: Generative model for handwritten digits}
\label{sec:mnist}
Next, we demonstrate the effectiveness of our Hamiltonian variational inference approach for learning deep generative neural network models. These models are fitted to a binarized version of the MNIST dataset as e.g. used in~\cite{Uria2013b}. This dataset consists of 70000 data vectors ${x_i}$, each of which represents a black-and-white image of a handwritten digit. The task of modelling the distribution of these handwritten digit images is often used as a comparative benchmark for probability density and mass modeling approaches.

Our generative model $p(x_{i},z_{i})$ consists of a spherical Gaussian prior $p(z_{i}) = \mathcal{N}(0,\mathbf{I})$, and conditional likelihood (or \emph{decoder}) $p_\theta(x_{i}|z_{i})$ parameterized with either a fully connected neural network as in~\cite{kingma2013auto,rezende2014stochastic}, or a convolutional network as in~\cite{dosovitskiy2014learning}. The network takes as input the latent variables $z_{i}$, and outputs the parameters of a conditionally independent (Bernoulli) distribution over the pixels.

Since we now have a dataset consisting of multiple datapoints ${x_i}$, with separate latent variables ${z_i}$ per datapoint, it is efficient to let the distribution $q(z|x)$ be an explicit function of the data $x_i$, since in that case there is often no necessity for 'local' variational parameters $\theta$ per individual datapoint $x_i$; instead, $q$ maps from global parameters $\theta$ and local observed value $x_i$ to a distribution over the local latent variable(s) $z_i$. We can then optimize over $\theta$ for all observations $x_i$ jointly. The joint lower bound to be optimized is given by
\[
\sum_{i=1}^{n} \log p(x_{i}) \geq \sum_{i=1}^{n} \E_{\qT(z_i|x_i)}[\log p(z_i,x_i) - \log \qT(z_i|x_i)],
\]
of which an unbiased estimator (and its gradients) can be constructed by sampling minibatches of data $x_i$ from the empirical distribution and sampling $z_i$ from $\qT(z_i|x_i$).

One flexible way of parameterizing the posterior approximation $\qT(z_{i}|x_{i})$ is by using an inference network as in Helmholtz machines \cite{hinton1994autoencoders} or the related variational auto-encoders (VAE)~\cite{kingma2013auto,rezende2014stochastic}.

We can augment or replace such inference networks with the MCMC variational approximations developed here, as the parameters $\theta$ of the Markov chain can also be shared over all data vectors $x_{i}$.


Specifically, we replace or augment inference networks as used in~\cite{kingma2013auto,rezende2014stochastic} with a Hamiltonian posterior approximation as described in Algorithm~\ref{algo:hmcvb}, with $T=1$ and a varying number of leapfrog steps. The auxiliary inference model $r(v|x,z)$ is chosen to be a fully-connected neural network with one deterministic hidden layer with $n_h=300$ hidden units with softplus ($\log(1+\exp(x))$) activations and a Gaussian output variable with diagonal covariance. We tested two variants of the distribution $q(z_0|x)$. In one case, we let this distribution be a Gaussian with a mean and diagonal covariance structure that are learned, but independent of the datapoint $x$. In the second case, we let $q(z_0|x)$ be an inference network like $r(v|x,z)$, with two layers of $n_h$ hidden units, softplus activations and Gaussian output with diagonal covariance structure. 

In a third experiment, we replaced the fully-connected networks with convolutional networks in both the inference model and the generative model. The inference model consists of three convolutional layers with 5$\times$5 filters, [16,32,32] feature maps, stride of 2 and softplus activations. The convolutional layers are followed by a single fully-connected layer with $n_h=300$ units and softplus activations. The architecture of the generative model mirrors the inference model but with stride replaced by upsampling, similar to~\cite{dosovitskiy2014learning}. The number of leapfrog steps was varied from 0 to 16. After broader model search with a validation set, we trained a final model with 16 leapfrog steps and $n_h=800$.

\begin{table}[t]
\caption{Comparison of our approach to other recent methods in the literature. We compare the average marginal log-likelihood measured in nats of the digits in the MNIST test set. See section~\ref{sec:mnist} for details.}
\begin{center}
\begin{tabular}{ l c c}
\textbf{Model}&$\log p(x)$&$\log p(x)$\\
 & $\leq -$ & $= -$ \\
\hline
\multicolumn{3}{l}{\rule{0pt}{3ex} \textbf{HVI + fully-connected VAE:}}\\
\textit{Without inference network:}\\
5 leapfrog steps & 90.86 & 87.16\\
10 leapfrog steps & 87.60 & 85.56\\
\textit{With inference network:}\\
No leapfrog steps & 94.18 & 88.95 \\
1 leapfrog step & 91.70 & 88.08\\
4 leapfrog steps & 89.82 & 86.40\\
8 leapfrog steps & 88.30 & 85.51\\
\hline
\multicolumn{3}{l}{\rule{0pt}{3ex} \textbf{HVI + convolutional VAE:}}\\
No leapfrog steps & 86.66 & 	83.20 \\
1 leapfrog step & 85.40 & 82.98 \\
2 leapfrog steps & 85.17 & 82.96 \\
4 leapfrog steps & 84.94 & 82.78 \\
8 leapfrog steps & 84.81 & 82.72 \\
16 leapfrog steps & 84.11 & 82.22 \\
16 leapfrog steps, $n_h = 800$ & 83.49 & 81.94 \\
\hline
\multicolumn{3}{l}{\rule{0pt}{3ex}\textbf{From~\cite{gregor2015draw}:}} \\
DBN 2hl && 84.55\\
EoNADE && 85.10\\
DARN 1hl & 88.30 & 84.13\\
DARN 12hl & 87.72 \\
DRAW & 80.97 & 
\end{tabular}
\end{center}
\label{table:mnist}\end{table}

Stochastic gradient-based optimization was performed using Adam~\cite{kingma2014adam} with default hyper-parameters. Before fitting our models to the full training set, the model hyper-parameters and number of training epochs were determined based on performance on a validaton set of about 15\% of the available training data.  The marginal likelihood of the test set was estimated with importance sampling by taking a Monte Carlo estimate of the expectation $p(x) = \E_{q(z|x)} [ p(x,z) / q(z|x) ]$ ~\cite{rezende2014stochastic} with over a thousand importance samples per test-set datapoint.

See table~\ref{table:mnist} for our numerical results and a comparison to reported results with other methods. Without an inference network and with 10 leapfrog steps we were able to achieve a mean test-set lower bound of $-87.6$, and an estimated mean marginal likelihood of $-85.56$. When no Hamiltonian dynamics was included the gap is more than 5 nats; the smaller difference of ~ 2 nats when 10 leapfrog steps were performed illustrates the bias-reduction effect of the MCMC chain. Our best result is $81.94$ nats with convolutional networks for inference and generation, and HVI with 16 leapfrog steps. This is slightly worse than the best reported number with DRAW~\cite{gregor2015draw}, a VAE with recurrent neural networks for both inference and generation. Our approaches are not mutually exclusive, and could indeed be combined for even better results.


\section{Specification of the Markov chain}
In addition to the core contributions presented above, we now present a more detailed analysis of some possible specifications of the Markov chain used in the variational approximation. We discuss the impact of different specification choices on the theoretical and practical performance of the algorithm.

\subsection{Detailed balance}
\label{sec:detail}
For practical MCMC inference we almost always use a transition operator that satisfies \textit{detailed balance}, i.e. a transition operator $q_{t}(z_{t}|x,z_{t-1})$ for which we have
\[
\frac{p(x,z_{t})\rev{q}_{t}(z_{t-1}|x,z_{t})}{p(x,z_{t-1})q_{t}(z_t|x,z_{t-1})} = 1,
\]
where $\rev{q}_t(z_{t-1}|x,z_{t})$ denotes $q_t(z_t|x,z_{t-1})$ with its $z$ arguments reversed (not $q(z_{t-1}|x,z_{t})$: the conditional pdf of $z_{t-1}$ given $z_t$ under $q$). If our transition operator satisfies detailed balance, we can divide $\alpha_t$ in Algorithm~\ref{algo:mcmclb} by the ratio above (i.e.\ 1) to give
\[
\log[\alpha_{t}] = \log r_t(z_{t-1}|x,z_{t}) - \log\rev{q}_{t}(z_{t-1}|x,z_{t}).
\]
By optimally choosing $r_t(z_{t-1}|x,z_{t})$ in this expression, we can make the expectation $\E_{q}\log[\alpha_{t}]$ non-negative: what is required is that $r_t()$ is a predictor of the reverse dynamics that is equal or better than $\rev{q}_t()$. If the iterate $z_{t-1}$ has converged to the posterior distribution $p(z|x)$ by running the Markov chain for a sufficient number of steps, then it follows from detailed balance that $\rev{q}_{t}(z_{t-1}|x,z_{t})=q(z_{t-1}|x,z_{t})$. In that case choosing $r_{t}(z_{t-1}|x,z_{t}) = \rev{q}_{t}(z_{t-1}|x,z_{t})$ is optimal, and the lower bound is unaffected by the transition. If, on the other hand, the chain has not fully \textit{mixed} yet, then $\rev{q}_{t}(z_{t-1}|x,z_{t}) \neq q(z_{t-1}|x,z_{t})$: the last iterate $z_{t-1}$ will then have a predictable dependence on the initial conditions which allows us to choose $r_t(z_{t-1}|x,z_{t})$ in such a way that $E_{q} \log[\alpha_{t}]$ is positive and improves our lower bound. Hence a stochastic transition respecting detailed balance always improves our variational posterior approximation unless it is already perfect! In practice, we can only use this to improve our auxiliary lower bound if we also have an adequately powerful model $r_t(z_{t-1}|x,z_t)$ that can be made sufficiently close to $q(z_{t-1}|x,z_{t})$.

A practical transition operator that satisfies detailed balance is Gibbs sampling, which can be trivially integrated into our framework as shown in Section~\ref{sec:gauss_gibbs}. Another popular way of ensuring our transitions satisfy detailed balance is by correcting them using Metropolis-Hastings rejection. In the latter case, the stochastic transition operator $q_{t}(z_{t}|x,z_{t-1})$ is constructed in two steps: First a stochastic \textit{proposal} $z_{t}'$ is generated from a distribution $\phi(z_{t}'|z_{t-1})$. Next, the \textit{acceptance probability} is calculated as
\begin{align*}
\rho(z_{t-1},z_{t}') = \text{min}\left[\frac{p(x,z_{t}')\phi(z_{t-1}|z_{t}')}{p(x,z_{t-1})\phi(z_{t}'|z_{t-1})}, 1\right].
\end{align*}
Finally, $z_{t}$ is set to $z_{t}'$ with probability $\rho(z_{t-1},z_{t}')$, and to $z_{t-1}$ with probability $1-\rho(z_{t-1},z_{t}')$. The density of the resulting stochastic transition operator $q_{t}(z_{t}|x,z_{t-1})$ cannot be calculated analytically since it involves an intractable integral over $\rho(z_{t-1},z_{t}')$. To incorporate a Metropolis-Hastings step into our variational objective we will thus need to explicitly represent the acceptance decision as an additional auxiliary binary random variable $a$. The Metropolis-Hastings step can then be interpreted as taking a reversible variable transformation with unit Jacobian:
\begin{align*}
z_{t-1} & \rightarrow \I[a=1]z_{t}' + \I[a=0]z_{t-1} \\
z_{t}' & \rightarrow \I[a=1]z_{t-1} + \I[a=0]z_{t}' \\
a & \rightarrow a.
\end{align*}
Evaluating our target density at the transformed variables, we get the following addition to the lower bound:
\begin{align*}
\log[\alpha_{t}] &= \log[p(x,z_t)/p(x,z_{t-1})] + \log[r_t(a|x,z_{t})] \\
&+ \I[a=1]\log[r_t(z_{t-1}|x,z_{t})] \\
&+ \I[a=0]\log[r_t(z_{t}'|x,z_{t})] \\
&- \log[q_t(z_{t}'|x,z_{t-1})q(a|z_{t}',z_{t-1},x)].
\end{align*}
Assuming we are working with a continuous variable $z$, the addition of the binary variable $a$ has the unfortunate effect that our Monte Carlo estimator of the lower bound is no longer a continuously differentiable function of the variational parameters $\theta$, which means we cannot use the gradient of the exact log posterior to form our gradient estimates. Estimators that do not use this gradient are available \cite{salimans2013fixed,paisley2012variational,ranganath2013black,mnih2014neural} but these typically have much higher variance. We can regain continuous differentiability with respect to $\theta$ by Rao-Blackwellizing our Monte Carlo lower bound approximation $L$ and calculating the expectation with respect to $q(a|z_{t}',z_{t-1},x)$ analytically. For short Markov chains this is indeed an attractive solution. For longer chains this strategy becomes computationally demanding as we need to do this for every step in the chain, thereby exploring all $2^{T}$ different paths created by the $T$ accept/reject decisions. Another good alternative is to simply omit the Metropolis-Hastings acceptance step from our transition operators and to rely on a flexible specification for $q()$ and $r()$ to sufficiently reduce any resulting bias.

\subsection{Annealed variational inference}
Annealed importance sampling is an MCMC strategy where the Markov chain consists of stochastic transitions $q_{t}(z_{t}|z_{t-1})$ that each satisfy detailed balance with respect to an unnormalized target distribution $\log[p_{t}(z)] = (1-\beta_{t})\log[q_{0}(z)] + \beta_{t}\log[p(x,z)]$, for $\beta_{t}$ gradually increasing from 0 to 1. The reverse model for annealed importance sampling is then constructed using transitions $r(z_{t-1}|z_t) = q_{t}(z_{t}|z_{t-1})p_{t}(z_{t-1})/p_{t}(z_{t})$, which are guaranteed to be normalized densities because of detailed balance. For this choice of posterior approximation and reverse model, the marginal likelihood lower bound is then given by
\[
\log p(x) \geq \E_{q} \sum_{t=1}^{T} (\beta_{t}-\beta_{t-1})\log[p(x,z_{t})/q_{0}(z_{t})].
\]
With $\beta_{0}=0, \beta_{T}=1$ this looks like the bound we have at $t=0$, but notice that the expectation is now taken with respect to a different distribution than $q_{0}$. Since this new approximation is strictly closer to $p(z|x)$ than the old approximation, its expectation of the log-ratio $\log[p(x,z_{t})/q_{0}(z_{t})]$ is strictly higher, and the lower bound will thus be improved.

The main advantage of annealed variational inference over other variational MCMC strategies is that it does not require explicit specification of the reverse model $r$, and that the addition of the Markov transitions to our base approximation $q_{0}(z)$ is guaranteed to improve the variational lower bound. A downside of using this scheme for variational inference is the requirement that the transitions $q(z_{t}|z_{t-1})$ satisfy detailed balance, which can be impractical for optimizing $q$.

\subsection{Using multiple iterates}
So far we have defined our variational approximation as the marginal of the last iterate in the Markov chain, i.e.\ $q(z_T|x)$. This is wasteful if our Markov chain consists of many steps, and practical MCMC algorithms therefore always use multiple samples $z_{T+1-K},\ldots,z_T$ from the Markov chain, with $K$ the number of samples. When using multiple samples obtained at different points in the Markov chain, our variational approximation effectively becomes a discrete \textit{mixture} over the marginals of the iterates that are used:
\begin{align*}
q(z|x) &= \frac{1}{K} \sum_{t = T+1-K}^{T} q(z_{t}|x) \nonumber\\
&= \sum_{t = T+1-K}^{T} \I(w=t)q(z_{t}|x),\\
&\text{ with } w \sim \text{Categorical}(T+1-K,\ldots,T).
\end{align*}
To use this mixture distribution to form our lower bound, we need to explicitly take into account the \textit{mixture indicator variable} $w$. This variable has a categorical distribution $q(w=t), t \in [T+1-K,\ldots,T]$ that puts equal probability on each of the $K$ last iterates of the Markov chain, the log of which is subtracted from our variational lower bound \eqref{eq:auxlowerbound}. This term is then offset by adding the corresponding log probability of that iterate under the inverse model $r(w=t|x,z)$.
The simplest specification for the inverse model is to set it equal to $q(w=t)$: In that case both terms cancel, and we're effectively just taking the average of the last $K$ lower bounds $L$ computed by Algorithm~\ref{algo:mcmclb}. Although suboptimal, we find this to be an effective method of reducing variance when working with longer Markov chains. An alternative, potentially more optimal approach would be to also specify the inverse model for $w$ using a flexible parametric function such as a neural network, taking $x$ and the sampled $z$ as inputs.

\subsection{Sequential MCVI}
In Algorithm~\ref{algo:mcvi} we suggest optimizing the bound over all MCMC steps jointly, which is expected to give the best results for a fixed number of MCMC steps. Another approach is to optimize the MCMC steps sequentially, by maximizing the local bound contributions $\E_{q} \log[\alpha_{t}]$. Using this approach, we can take any existing variational approximation and improve it by adding one or more MCMC steps. Improving an existing approximation in this way gives us an easier optimization problem, and can be compared to how boosting algorithms are used to iteratively fit regression models. 
\begin{algorithm}[H]
\caption{Sequential MCVI}
\label{algo:bmcvi}
\begin{algorithmic}
\REQUIRE Unnormalized log posterior $\log p(x,z)$
\REQUIRE Variational approximation $q(z_0|x)$
\FOR{$t = 1:T$}
\STATE Add transition operator $q_{t}(z_t|x,z_{t-1})$ and inverse model $r_t(z_{t-1}|x,z_t)$.
\STATE Choose the new parameters by maximizing the local lower bound contribution $\E_{q(z_{t},z_{t-1})} \log[\alpha_{t}]$
\STATE Set the new posterior approximation equal to $q(z_t|x) = \int q_t(z_t|x,z_{t-1})q(z_{t-1}|x) dz_{t-1}$
\ENDFOR
\RETURN the final posterior approximation $q(z_{T}|x)$
\end{algorithmic}
\end{algorithm}

\section{Conclusion}
By using auxiliary variables in combination with stochastic gradient variational inference we can construct posterior approximations that are much better than can be obtained using only simpler exponential family forms. One way of improving variational inference is by integrating one or more MCMC steps into the approximation. By doing so we can bridge the accuracy/speed gap between MCMC and variational inference and get the best of both worlds. 

\bibliographystyle{icml2015}
\bibliography{biball}

\end{document}